\documentclass[useAMS,usenatbib]{mn2e}
\input{epsf}
\usepackage{amssymb}
\usepackage{natbib}
\usepackage{color}
\usepackage{journals}
\usepackage{graphicx}

\newbox\grsign \setbox\grsign=\hbox{$>$} \newdimen\grdimen \grdimen=\ht\grsign
\newbox\simlessbox \newbox\simgreatbox
\setbox\simgreatbox=\hbox{\raise.5ex\hbox{$>$}\llap
     {\lower.5ex\hbox{$\sim$}}}\ht1=\grdimen\dp1=0pt
\setbox\simlessbox=\hbox{\raise.5ex\hbox{$<$}\llap 
     {\lower.5ex\hbox{$\sim$}}}\ht2=\grdimen\dp2=0pt

\newcommand{\hMpc}{{\ifmmode{h^{-1}{\rm Mpc}}\else{$h^{-1}$Mpc }\fi}}
\newcommand{\hGpc}{{\ifmmode{h^{-1}{\rm Gpc}}\else{$h^{-1}$Gpc }\fi}}
\newcommand{\hkpc}{{\ifmmode{h^{-1}{\rm kpc}}\else{$h^{-1}$kpc }\fi}}
\newcommand{\hMsun}{{\ifmmode{h^{-1}{\rm {M_{\odot}}}}\else{$h^{-1}{\rm{M_{\odot}}}$}\fi}}
\newcommand{\Msun}{{\ifmmode{{\rm {M_{\odot}}}}\else{${\rm{M_{\odot}}}$}\fi}}

\title[Tension in the SN sector of $H_{0}$ measurement]{Intrinsic tension in the supernova sector of the local Hubble constant measurement and its implications}
\author[R. Wojtak]{Rados{\l}aw~Wojtak$^{1}$\thanks{E-mail: radek.wojtak@nbi.ku.dk} \& Jens Hjorth$^{1}$ \\
$^{1}$DARK, Niels Bohr Institute, University of Copenhagen, Jagtvej 128, 2200 Copenhagen, Denmark  \\
}
\begin{document}

\maketitle

\begin{abstract}
We reanalyse observations of type Ia supernovae (SNe) and Cepheids used in the local determination of the Hubble constant and
find strong evidence that SN standardisation in the calibration sample (galaxies with observed Cepheids) require a steeper slope of the colour correction than in the cosmological sample
(galaxies in the Hubble flow). The colour correction in the calibration sample
is consistent with being entirely due to an extinction correction due to dust with properties similar to that of the Milky Way 
($R_B\approx 4.6\pm0.4$) and there is no evidence for intrinsic scatter in the SN peak magnitudes. An immediate consequence of this finding 
is that the local measurement of the Hubble constant becomes dependent on the choice of SN reference colour, i.e., the colour of an 
unreddened SN. Specifically, the Hubble constant inferred 
from the same observations decreases gradually with the reference colour assumed in the SN standardisation. We recover the Hubble constant measured 
by SH0ES for the standard choice of reference colour (SALT2 colour parameter $c=0$) while 
for a reference colour which coincides with the blue end of the observed SN colour distribution ($c\approx-0.13$), the Hubble constant from {\em Planck} observations of the CMB (assuming a flat $\Lambda$CDM cosmological model) is recovered. 
These results are intriguing in that they may provide an avenue for
resolving the Hubble tension. However, since there is no obvious physical basis for the differences in colour corrections in the two SN samples, the origin of these
require further investigations.
%of possible systematic biases and 
%improved understanding of the physics behind the colour correction and possible
%residual scatter in SN Hubble diagrams.

\end{abstract}

\begin{keywords}
cosmology: observations, distance scale, cosmological parameters -- methods: statistical
\end{keywords}

\section{Introduction}

The discrepancy between the Hubble constant measured from observations of type Ia supernovae (SNe) and Cepheids 
with geometric distance calibrations \citep{Riess2019,Riess2021,Riess2022}, and from {\em Planck} observations of the CMB assuming 
a flat $\Lambda$CDM cosmological model \citep{Planck2020_cosmo} has recently drawn much attention. With $H_{0}=73.04\pm1.04$~km~s$^{-1}$~Mpc$^{-1}$ derived from the most recent advances in Cepheid and SN observations of 
the SH0ES collaboration \citep[for Supernova H0 for the Equation of State][]{Riess2022} and $H_{0}=67.36\pm0.54$~km~s$^{-1}$~Mpc$^{-1}$ inferred from the {\em Planck} data \citep{Planck2020_cosmo}, it is currently the strongest divergence in cosmological measurements with a confidence level reaching $5\sigma$. The nature of this tension 
is unknown and its origin may lie in either unaccounted systematic effects or an incomplete 
theoretical framework of the standard $\Lambda$CDM cosmological model.

Multiple extensions or modifications of the standard $\Lambda$CDM cosmological model were recently proposed as potential 
solutions to the Hubble constant tension \citep{DiVal2021}. However, none of the current proposals give a satisfying solution 
without affecting a wide range of other cosmological measurements. Arguably the most promising scenario involves 
early dark energy which is a hypothesised extra energy component manifesting itself before photon decoupling 
\citep{Poulin2019}. This model is theoretically designed to shorten the sound horizon scale and thus elevate the Hubble constant value derived from the CMB, while keeping the observed baryon acoustic oscillations (BAO) angular scale unchanged \citep{Knox2020}. The main drawback of this proposal is that the {\em Planck} data do not provide any statistically significant evidence for early dark energy \citep{Arendse2020,Fondi2022,Vagnozzi2021}. 
Furthermore, early dark energy also has been proven to affect the power spectrum derived from the CMB in a way that spoils a fair consistency between 
the {\em Planck} cosmology and constraints from observations probing large scale structure \citep{Hill2020}. At the opposite end 
of the spectrum of cosmological scenarios lie models which attempt to reconcile the local and {\em Planck} measurements of the Hubble constant 
by {\em ad hoc} modifications of the very recent expansion history by means of tuning a time-dependent equation of state for dark 
energy \citep{DiVal2021}. This approach, however, ignores that fact the Hubble constant measured locally is not a direct observable 
which can be used as a prior in cosmological analyses with type Ia SN data. Finding an expansion history which interpolates between the 
local value of the Hubble constant and the {\em Planck} cosmology at high redshifts does not resolve the Hubble constant tension, but rather
shifts the problem of discordant distance scales to a discrepancy between the absolute luminosity of type Ia SNe 
calibrated with Cepheids and its analog obtained in the so-called inverse distance ladder method based on distance scales calibrated 
with the {\em Planck} data \citep{Efs2021,Cam2021}.

Independent measurements of the Hubble constant on intermediate cosmic distance scales can potentially provide decisive arguments supporting or ruling out the interpretation of the Hubble constant tension as a cosmological anomaly. Arguably the best technique operating on these scales is using time delays of gravitationally lensed and multiply imaged variable sources. Although substantial progress has been made in this field, present estimates of the Hubble constant are limited by the accuracy of lens models and range between the SH0ES \citep{Wong2020} 
and {\em Planck} values \citep{Birrer2020}.

Despite a growing conviction that the Hubble constant tension is a cosmological anomaly, alternative scenarios involving hidden 
and currently unaccounted systematic effects are not completely ruled out. The majority of studies in this area were undertaken 
to test the robustness of the local measurement of the Hubble constant with respect to possible changes in modelling 
the Cepheid data. A wide range of possible systematic effects including non-standard colour correction \citep{Mor2021}, 
the impact of outliers \citep{Efs2014}, blending and many other effects \citep[for an exhausting list of tests see][]{Riess2022} 
were shown to have a negligible impact on the Hubble constant determination \citep{Riess2022}. The ultimate robustness test of the 
Cepheid data sector should involve an alternative distance calibration which could replace entirely the Cepheid rod 
of the distance ladder. Cepheid-independent measurements of the Hubble constant were carried out recently following 
advances in calibrating cosmological distances with the tip of red giant branch \citep[TRGB;][]{Freed2019} 
or surface brightness fluctuations \citep[SBF;][]{Khetan2021,Garnavich2022}. The results are broadly consistent 
with the Hubble constant value inferred from Cepheid calibration. However, none of the measurements are currently precise 
enough to discriminate decisively between the {\em Planck} and Cepheid-based values of the Hubble constant.

Compared to quite exhausting robustness tests of the Cepheid sector, rather little attention has been drawn to type Ia SNe 
as a possible source of unknown systematic effects in the local $H_{0}$ determination. This is 
worth pursuing considering the purely phenomenological nature of the model used to standardise SN peak magnitudes for distance 
measurements \citep{Tripp1998}. Moreover, the fact that the apparent intrinsic scatter of $\approx 0.1$~mag 
in the Hubble diagram with type Ia SNe found consistently in all independent studies \citep[see e.g.][]{Scolnic2018,Jones2019} 
is comparable to the difference between the local and {\em Planck} values of the Hubble constant expressed in distance 
moduli, i.e., $\Delta\mu\approx0.16$, is intriguing. The robustness of the Hubble constant measurement based on observations of 
type Ia SNe directly relies on accurate distance propagation between SN host galaxies with observed Cepheids (calibration sample) and SN 
host galaxies in the Hubble flow (cosmological sample). If the currently used SN standardisation is not equally accurate in both SN samples, biases may potentially affect the Hubble constant measurement.

The main goal of this study is to quantify to what extent the calibration and cosmological 
SN samples are consistent with the same universal colour correction which, alongside the correction 
due to differences in light curve shape \citep{Phillips1993}, constitutes the commonly used SN standardisation model \citep{Tripp1998}. Our study 
involves reanalysis of existing observations 
of Cepheids and type Ia SNe employing observationally motivated extensions to the standard approach adopted in \citet{Riess2022}. 
Based on a revised SN colour correction resulting from our analysis, we rederive the Hubble constant and discuss the conditions 
for resolving the Hubble constant tension.

The outline of the paper is as follows. In section 2 we describe the data, models and methods used in our study. The main 
results are presented in section 3. This includes detection of an anomaly in the SN data sector of the local Hubble constant 
determination (section 3.1), the evidence for discrepant colour corrections in the calibration and cosmological SN samples (section 3.2) and its 
impact on the Hubble constant determination (section 3.3). We discuss the results in section 4 and summarise our findings in section 5.

\section{Data and model}

We use the complete data set which was the basis for the recent measurements of 
the Hubble constant presented in \citet{Riess2019} and \citet{Riess2021}. The data 
comprise observations of Cepheids and type Ia SNe, as well as a range 
of geometric distance estimates. For the sake of better representation of the data structure 
in relation to the model, we split the data into seven independent blocks, each described 
by its own likelihood and the corresponding set of parameters. Table~\ref{table-likelihood} 
provides a concise description of each data block in terms of data, likelihood formula and model parameters. 

\subsection{Data blocks}

The first three data blocks listed in Table~\ref{table-likelihood} comprise measurements 
of reddening-free Wesenheit apparent magnitudes, $m_{\rm F160W}^{W}$, and pulsation periods, 
$P$, of Cepheids observed in the Milky Way \citep[$L_{\rm MW}$;][]{Riess2021}, the Large Magellanic 
Cloud \citep[$L_{\rm LMC}$;][]{Riess2019} and 20 galaxies \citep[$L_{\rm cal}$;][]{Riess2016} including 19 type 
Ia SNe host galaxies and the megamaser galaxy NGC 4258. These measurements constrain 
distance moduli $\mu$ via the following equation:
\begin{equation}
m_{\rm F160W}^{W}=\mu+M_{\rm F160W}+b_{W}(\log_{10}P-1)+z_{W}\Delta[O/H],
\label{Cep}
\end{equation}
where the {\em Hubble Space Telescope} ${\rm F160W}$-band absolute magnitude, $M_{\rm F160W}$, and the free coefficients $b_{W}$ and $z_{W}$ are measured 
directly from the data. The additional term in eq.~(\ref{Cep}) incorporates corrections due to 
metallicity $\Delta[\textrm{O/H}]$ which is directly measured for all Milky Way Cepheids and local environments 
of Cepheids in galaxies of the calibration sample. Following \citet{Riess2019} we assume that 
all Cepheids observed in the LMC have the same metallicity equal to the mean $\Delta[\textrm{O/H}]=-0.30$~dex
found in the LMC. For metallicity $[O/H]$ measurements provided by \citet{Riess2016} we derive 
$\Delta[\textrm{O/H}]$ assuming the solar metallicity given by the calibration of these metallicity 
estimates, i.e., $8.93$ \citep{Anders1989}. Wesenheit magnitudes of Cepheids 
in the 20 calibration galaxies are calculated from magnitudes $m_{\rm F160W}$ and colours 
$m_{\rm F555W}-m_{\rm F814W}$ provided by \citet{Riess2016} using the relation
\begin{equation}
m_{\rm F160W}^{W}=m_{\rm F160W}+R(m_{\rm F555W}-m_{\rm F814W})
\end{equation}
with $R=-0.386$ corresponding to an extinction coefficient $R_{\rm B}=4.3$ in the reddening law of \citet{Fitzpatrick1999}, 
as assumed in \citet{Riess2019} and \citet{Riess2021}.

Distances to Cepheids in the Milky Way are constrained by {\em Gaia} measurements of their parallaxes, 
$\pi$, \citep{Gaia2021}. Following \citet{Riess2021} we include zero point $zp$ as a free parameter in order to obtain 
an unbiased relation between observed parallaxes, $\pi$, and distance moduli using the following equation
\begin{equation}
\pi+zp=10^{-0.2(\mu-10)}.
\end{equation}
As described explicitly in Table~\ref{table-likelihood}, the likelihood for the Milky Way Cepheids accounts for 
uncertainties in the parallax and magnitude measurements. Here we also include an error of $0.01$~dex 
in the metallicities. The errors in likelihoods $L_{\rm LMC}$ and $L_{\rm cal}$ are given by 
the measurement uncertainties in $m_{\rm F160W}^{W}$ provided by \citet{Riess2016} and \citet{Riess2019}. Distance moduli 
of the LMC and the 20 calibrator galaxies are described by latent variables for which constraints are obtained 
as a byproduct of the entire analysis combining all likelihoods.

Independent measurements of geometric distances to the LMC from detached eclipsing binaries 
\citep{Pietrz2019} and to NGC 4258 from megamasers \citep{Reid2019} are included in likelihoods $L_{\rm LMC}$ 
and $L_{\rm cal}$. Together with parallax distances to Cepheids in the Milky Way, these measurements serve 
as the anchors of the cosmological distance scale entering the Hubble constant determination.

The last two data blocks in Table~\ref{table-likelihood} include observations of type Ia SNe. We use light 
curve parameters obtained by \citet{Scolnic2015} using the SALT2 fitting methodology \citep{Guy2005,Scolnic2016}. 
Distance moduli are derived from peak apparent magnitudes, $m_{B}$, by applying corrections related to the 
light curve shape quantified by $x_{1}$ and colour parameter $c$, where the latter is thought to combine effectively two physical effects: extinction correction due to intervening dust in SN host galaxies and a possible correlation between the absolute magnitude and the SN intrinsic colour. We employ the standard correction model formulated 
by \citet{Tripp1998},
\begin{equation}
m_{B}=\mu+M_{B}-\alpha x_{1}+\beta c,
\label{SN-cal}
\end{equation}
where B-band absolute magnitude, $M_{B}$, and correction coefficients $\alpha$ and $\beta$ are measured directly from 
the SN data. Errors in distance moduli adopted in the SN likelihoods are given explicitly in 
Table~\ref{table-likelihood}. They include all elements of covariance matrices obtained for individual SNe, 
intrinsic scatter $\sigma_{\rm int}$ and extra error due to unconstrained peculiar velocities (included only in $L_{\rm SN}$) 
with $c\sigma_{z}=200$~km~s${^{-1}}$ \citep{Carrick2015}. 

For the cosmological data block ($L_{\rm SN}$) we selected SNe using the same criteria as those adopted by 
\citet{Riess2016}. The primary selection condition is given by redshift range $0.023<z<0.15$ and cuts in light curve 
parameters: $|c|<0.3$ and $|x_{1}|<3$. In addition, SNe with low quality fits are rejected. SNe passing 
the fit quality check (223) are characterised by sufficient goodness of fit (${\rm fitprob}>0.001$) and well-constrained 
light-curve parameters with errors $<1.5$ for $x_{1}$, $<2$~days for the peak time and $<0.2$~mag for the corrected peak  magnitude.

Finally, as a novel approach, in this study we will allow for SNe in the calibration block to be characterised by an independent 
slope, $\beta_{\rm cal}$, of the colour correction, i.e.,
\begin{equation}
m_{B}=\mu+M_{B}-\alpha x_{1}+\beta_{\rm cal} c,
\label{SN}
\end{equation}
and associated intrinsic scatter, $\sigma_{\rm int\,cal}$, in $M_B$. These extra parameters will enable us to test the standard approach of
assuming universality of the colour correction 
($\beta_{\rm cal}\equiv\beta$) and intrinsic scatter ($\sigma_{\rm int}\equiv\sigma_{\rm int\,cal}$) 
and explore its impact on the local $H_{0}$ determination.

\begin{table*}
\begin{center}
\begin{tabular}{llcl} Label
 & Data & $\ln L$ & Parameters \\
  \hline
   \hline
 & Cepheids & & \\
 \hline
 \hline
LMC & $\{m_{\rm F160W}^{W},P,\sigma_{m_{\rm F160W}^{W}}\}^{a}$ & $-\frac{1}{2}\sum\frac{(m_{\rm F160W\,i}^{W}-\hat{\mu}_{\rm LMC}-M_{\rm F160W}^{W}-b_{W}(\log_{10}P_{i}-1)-z_{W}\Delta[O/H])^{2}}{\sigma^{2}_{m_{\rm F160W}^{W}\,i}}$ & $M_{\rm F160W}^{W},b_{W},z_{W},$\\
 & $\Delta[O/H]=-0.3$ & & $\hat{\mu}_{\rm LMC}$\\
\hline
MW & $\{\pi,\sigma_{\pi},m_{\rm F160W}^{W},\sigma_{m_{\rm F160W}^{W}},P,$ & $-\frac{1}{2}\sum\big([\pi_{i}+zp-10^{-0.2(\mu_{i}-10)}]^{2}/\sigma_{i}^{2}+\ln\sigma_{i}^{2}\big)$ & $M_{\rm F160W}^{W},b_{W},z_{W},$ \\
  &  & $\mu_{i}=m_{\rm F160W\,i}^{W}-M_{\rm F160W}^{W}-b_{W}(\log_{10}P_{i}-1)-z_{W}\Delta[O/H]_{i}$ & $zp$\\
 & $\Delta[O/H]\}^{b,c},\sigma_{\Delta[O/H]} =0.01$ & $\sigma_{i}^{2}=\sigma_{\pi\;i}^{2}+(\ln(10)/5)^{2}10^{-0.4(\mu_{i}-10)}\big(\sigma^{2}_{m_{\rm F160W}^{W}\,i}+z_{W}^{2}\sigma^{2}_{\Delta[O/H]}\big)$  & \\
 \hline
cal & 
$\{\{m_{\rm F160W}^{W},\sigma_{m_{\rm F160W}^{W}},P\},$ & 
$-\frac{1}{2}\sum_{i=1}^{20}\sum_{j}\frac{(m_{\rm F160W\,ij}^{W}-\hat{\mu}_{i}-M_{\rm F160W}^{W}-b_{W}(\log_{10}P_{ij}-1)-z_{W}\Delta[O/H]_{i})^{2}}{\sigma^{2}_{m_{\rm F160W}^{W}\,ij}}$ & $M_{\rm F160W}^{W},b_{W},z_{W},$ \\
 & $\Delta[O/H]\}^{d}$ &  & $\hat{\mu}_{1},...,\hat{\mu}_{20}$ \\
 \hline
  \hline
  & Anchors & & \\
 \hline
  \hline
4258 & $\mu_{4258}=29.398^{e}$ & $ -\frac{1}{2}\frac{(\mu_{4258}-\hat{\mu}_{4258})^{2}}{\sigma_{\mu}^{2}} $ & $\hat{\mu}_{4258}\equiv\hat{\mu}_{20}$ \\
 & $\sigma_{\mu}=0.032$ & &  \\
 \hline
${\rm LMC\,dist}$ & $\mu_{\rm LMC}=18.477^{f}$ & $ -\frac{1}{2}\frac{(\mu_{\rm LMC}-\hat{\mu}_{\rm LMC})^{2}}{\sigma_{\mu}^{2}} $ & $\hat{\mu}_{\rm LMC}$ \\
 & $\sigma_{\mu}=0.0263$ & &  \\
 \hline
  \hline
    & SNe & & \\
   \hline
 \hline
 ${\rm SN\,cal}$ & $\{m_{B},x_{1},c,{\rm COV}\}^{f}$ & 
 $-\frac{1}{2}\sum_{i=1}^{19}\big(\frac{(m_{B\,i}-\hat{\mu}_{i}-M_{B}+\alpha x_{1}-\beta_{\rm cal} c)^{2}}{\sigma^{2}_{i}}+\ln\sigma_{i}^{2}\big)$ 
 & $M_{B},\alpha,\beta_{\rm cal},\sigma_{\rm int\,cal},$ \\
  & & $\sigma^{2}_{i}=\sigma_{m_{B}i}^{2}+\alpha^{2}\sigma_{x_{1}i}^{2}+\beta_{\rm cal}^{2}\sigma_{c\,i}^{2}+\sigma_{\rm int\,cal}^{2}$ & $\hat{\mu}_{1},...,\hat{\mu}_{19}$ \\
 & & $-2\beta_{\rm cal}{\rm COV}_{m_{B},c}+2\alpha{\rm COV}_{m_{B},x_{1}}-2\alpha\beta_{\rm cal}{\rm COV}_{c,x_{1}}$ &  \\
 \hline
SN & $\{m_{B},x_{1},c,{\rm COV},z\}^{g}$ & 
 $-\frac{1}{2}\sum\big(\frac{(m_{B\,i}-\mu_{\rm Planck}(z_{i})+5\log_{10}(H_{0}/H_{0,{\rm Planck}})-M_{B}+\alpha x_{1}-\beta c)^{2}}{\sigma^{2}_{i}}+\ln\sigma_{i}^{2}\big)$ 
 & $M_{B},\alpha,\beta,\sigma_{\rm int},H_{0}$ \\
  & & $\sigma^{2}_{i}=\sigma_{m_{B}i}^{2}+\alpha^{2}\sigma_{x_{1}i}^{2}+\beta^{2}\sigma_{c\,i}^{2}+\sigma_{\rm int}^{2}+(5/\ln10)^{2}(\sigma_{z}/z)^{2} $ &  \\
 & $c\sigma_{z}=200\,{\rm km}\,{\rm s}^{-1}$ & $-2\beta{\rm COV}_{m_{B},c}+2\alpha{\rm COV}_{m_{B},x_{1}}-2\alpha\beta{\rm COV}_{c,x_{1}}$ & 
\end{tabular}
\caption{Summary of data sets, likelihood functions and parameters used in this study. The Cepheid data block (Cepheids) comprises 
observations of Cepheids in the Large Magellanic Cloud (${\rm LMC}$), the Milky Way (${\rm MW}$) and 20 galaxies including 
19 SN host galaxies and the NGC 4258 megamaser galaxy (${\rm cal}$). The distance anchor data block (Anchors) contains measurements 
of geometric distances to the LMC and NGC 4258. The SN data block (SNe) comprises light curve parameters of type Ia 
SNe in the 19 calibration galaxies (${\rm SN\,cal}$) and in the cosmological sample at redshifts $0.023<z<0.15$ (${\rm SN}$). The notation 
uses $\sigma_{\rm x}$ to denote measurement uncertainty for quantity x and $\rm COV_{\rm x,y}$ to denote the element of a covariance matrix for 
variables $x$ and $y$. The model adopted in this study involves 11 primary parameters: ($M_{\rm F160W},b_{W},z_{W}$) describing Cepheid calibration 
given by eq.~(\ref{Cep}), zero point $zp$ of the Cepheid parallaxes from {\it GAIA}, ($M_{B},\alpha,\beta,\beta_{\rm cal}$) describing type Ia SN 
calibration given by eq.~(\ref{SN}) (where $\beta_{\rm cal}$ refers to the calibration sample), ($\sigma_{\rm int},\sigma_{\rm int\,cal}$) describing intrinsic scatter in the corrected magnitudes of SNe respectively in the cosmological and calibration samples, and the Hubble constant $H_{0}$. In addition, the model includes 21 latent parameters which quantify the
distance moduli of 19 calibration galaxies ($\mu_{1},...,\mu_{19}$), the NGC 4258 megamaser galaxy ($\mu_{20}\equiv\mu_{4258}$) and the LMC ($\mu_{\rm LMC}$).\\
References for data sources: $^{a}$\citet{Riess2019}, $^{b}$\citet{Riess2021},$^{c}$\citet{Gaia2021}, $^{d}$\citet{Riess2016}, $^{e}$\citet{Reid2019}, $^{f}$\citet{Pietrz2019},
$^{g}$\citet{Scolnic2015}
}
\label{table-likelihood}
\end{center}
\end{table*}

\subsection{Parameter estimation}
We constrain model parameters using the likelihood function which is a product of likelihoods from all seven data blocks, i.e.,
\begin{equation}
L\propto L_{\rm LMC}\times L_{\rm MW}\times L_{\rm cal}\times L_{\rm 4258}\times L_{\rm LMC\,dist}\times L_{\rm SN\,cal}\times L_{\rm SN}.
\end{equation}
Our model is described by 11 primary parameters $(H_{0},M_{\rm F160W},b_{W},z_{W},z_{p},M_{B},\alpha,\beta,\sigma_{\rm int},\beta_{\rm cal}, \sigma_{\rm int\,cal})$ out of which two ($\beta_{\rm cal}$ and $\sigma_{\rm int\,cal}$ -- the colour correction slope and intrinsic scatter 
in the SN calibration sample) are optional. In addition, 21 latent variables are constrained as a byproduct of fitting. These are the
distance moduli of the 19 calibration galaxies, the LMC and the megamaser galaxy NGC 4258. Including these latent variable as extra nuisance 
parameters in the analysis is not relevant for constraining the primary parameters and one can alternatively employ a compressed 
likelihood obtained by analytical integration of $L$ over all latent variables. However, having access to the latent variables enables a range 
of sanity tests aimed at checking consistencies between different data blocks given the best fit model. In particular, in our study we scrutinise the
intrinsic consistency between the calibration ($L_{\rm SN\,cal}$) and cosmological ($L_{\rm SN}$) SN data blocks.

We use a Markov Chain Monte Carlo technique to integrate the posterior probability function and find best fit model parameters. The chains 
are computed with the \textit{emcee} code \citep{emcee}. Unless explicitly stated, best-fit parameters are provided as the posterior mean values 
and errors given by 16th and 84th percentiles of the marginalised probability distributions. The $1\sigma$ and $2\sigma$ confidence contours 
shown in all figures contain 68 and 95 per cent, respectively, of the corresponding 2-dimensional marginalised probability distributions.

The Hubble constant is derived assuming that the shape of the distance modulus as a function of redshift, which is independent of $H_{0}$, is given by the {\em Planck} cosmological 
model \citep{Planck2020_cosmo}. The respective formula for distance modulus reads
\begin{equation}
    \mu(z)=\mu_{\rm Planck}(z)+5\log_{10}(H_{\rm 0\,Planck}/H_{0}),
\end{equation}
where all quantities with Planck subscript are given by the {\em Planck} model while 
$H_{0}$ is a free parameter in our analysis (see also the SN block in Table~\ref{table-likelihood}). For the adopted SN redshift range in the cosmological data block, this strategy is in practice equivalent to the commonly used third-order approximation of cosmological distance as a function of redshift with fixed values of the 
deceleration parameter $q_{0}$ and the jerk parameter $j_{0}$ \citep{Visser2004}. Small variation in $q_{0}$ and $j_{0}$ within a wide range of dark matter 
and dark energy density parameters have been shown to have a negligible impact on the current estimation of the Hubble constant \citep{Riess2022}.

\section{Analysis}

\subsection{Baseline model}

\begin{figure*}
	\centering
	\includegraphics[width=\linewidth]{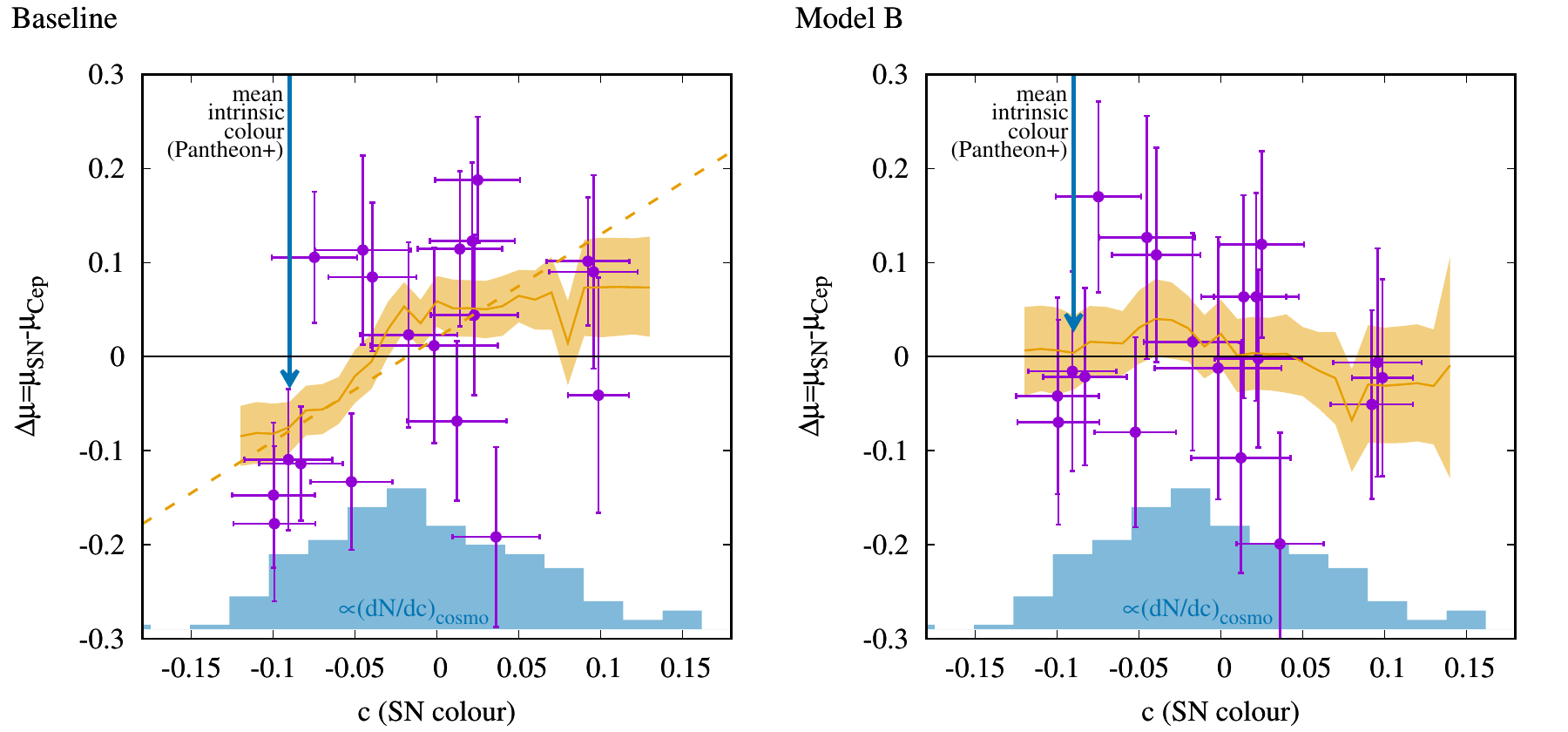}
	\caption{Comparison between distance moduli of 19 calibration galaxies obtained from type Ia SN light curve parameters 
	($\mu_{\rm SN}$) or from Cepheid observations calibrated with independent geometric distance measurements ($\mu_{\rm Cep}$), 
	as a function of SN colour parameter. Distance moduli are derived from fitting Cepheid and SN observations assuming 
	the baseline model with a universal SN colour correction (left) or model B (the most preferred among extensions to the baseline model considered in this study) with independent SN colour corrections in the 
	calibration and cosmological sample (right). The orange bands show the best fit $\Delta\mu$ (mean and $1\sigma$ range) 
	obtained in a rolling window with a width of $0.04$ and the dashed line in the left panel shows the best fit linear model. \textit{The left panel 
	demonstrates an anomalous trend signifying a biased colour correction in the calibration sample for the baseline model.}
	The right panel shows that model B with an independent SN colour correction in the calibration sample brings about consistency between the SN and Cepheid 
	observations. The blue arrow marks the current estimate of the mean SN intrinsic colour derived from the
	Pantheon+ SN sample assuming an exponential distribution of dust reddening \citep{Popovic2021}. All SNe in the calibration sample appear redder than the mean intrinsic colour suggesting that the 
	anomaly results most likely from insufficient correction for dust extinction in the baseline model. The histogram shows the distribution 
	of colour parameter $c$ in the cosmological sample.	
	}
	\label{19-res-c}
\end{figure*}

To validate our approach, we begin by fitting a model which resembles closely the fitting strategy adopted by \citet{Riess2016} or the baseline model in \citet{Riess2022}. 
The model includes the minimum number of primary parameters which are necessary to fit the Cepheid data ($M_{\rm F160W},b_{W},z_{W},z_{p}$) and 
SN data ($M_{B},\alpha,\beta,\sigma_{\rm int},H_{0}$). Here we assume that the colour correction coefficient and intrinsic scatter of SN corrected magnitudes are the same in the SN calibration ($L_{\rm SN\,cal}$) and cosmological ($L_{\rm SN}$) data blocks, i.e., $\beta=\beta_{\rm cal}$ and $\sigma_{\rm int}=\sigma_{\rm int\,cal}$.

Our best fit baseline model recovers all essential results obtained for the same data compilation in the original studies. In particular, 
we find excellent agreement with the Cepheid parameters $(M_{\rm F160W},b_{W},z_{W})$ measured by \citet{Riess2022},
the zero point $z_{p}$ 
of the Cepheid parallaxes from {\em Gaia} determined by \citet{Riess2021} and the SN calibration 
parameters $(M_{B},\alpha,\beta,\sigma_{\rm int})$ 
found for similar low-redshift $z\lesssim 0.1$ SN samples \citep[e.g.][]{Jones2019}. The best fit Hubble constant 
$H_{0}=(73.14\pm1.31)\,{\rm km}\,{\rm s}^{-1}\,{\rm Mpc}^{-1}$ is fully consistent with the original measurements \citep{Riess2019,Riess2021} based on the same calibration sample and the most recent updates in distance anchors as summarised in \citet{Riess2022}. Our measurement is at a $4\sigma$ discrepancy with the Hubble constant determination 
from {\em Planck} observations assuming a flat $\Lambda$CDM cosmological model \citep{Planck2020_cosmo}. This tension increases 
to a $5\sigma$ level in \cite{Riess2022} primarily due to $\sim 30$ per cent smaller errors resulting from the twice as large calibration sample.

\begin{table*}
\begin{center}
\begin{tabular}{lcccc}
\hline
 & baseline & model A & \textbf{model B}  & model C \\ 
 & & & \textbf{(most favoured)} & \\
 \hline
 SN parameters & & & \\
 \hline
 $\beta$ & $ 3.109 ^{+ 0.112 }_{- 0.112 }$ & $ 3.060 ^{+ 0.114 }_{- 0.115 }$ &  $\mathbf{3.057 ^{+ 0.114 }_{- 0.113 }}$ & $ 3.062 ^{+ 0.114 }_{- 0.113 }$ \\
 & & & & \\
$\sigma_{\rm int}$ & $ 0.115 ^{+ 0.008 }_{- 0.008 }$ & $ 0.115 ^{+ 0.008 }_{- 0.008 }$ & $ \mathbf{0.115 ^{+ 0.008 }_{- 0.008 }}$ & $ 0.113 ^{+ 0.008 }_{- 0.008 }$\\
 & & & & \\
$\beta_{\rm cal}$ & $\equiv\beta$ & $ 4.285 ^{+ 0.504 }_{- 0.503 }$ & $\mathbf{ 4.565 ^{+ 0.384 }_{- 0.379 }}$ & $ 4.110 ^{+ 0.542 }_{- 0.550 }$\\
 & & & & \\
$\sigma_{\rm int\,cal}$  & $\equiv\sigma_{\rm int}$ & $ <0.097$ & $\mathbf{\equiv 0}$ & $\equiv\sigma_{\rm int}$\\
 & & & & \\
 $M_{B}$ & $ -19.245 ^{+ 0.038 }_{- 0.037 }$ & $ -19.223 ^{+ 0.037 }_{- 0.037 }$ & $\mathbf{-19.216 ^{+ 0.032 }_{- 0.032 }}$& $ -19.228 ^{+ 0.040 }_{- 0.040 }$ \\
 & & & & \\
$\alpha$ & $ 0.129 ^{+ 0.008 }_{- 0.009 }$ & $ 0.129 ^{+ 0.008 }_{- 0.008 }$ & $ \mathbf{0.130 ^{+ 0.008 }_{- 0.008 }}$ & $ 0.129 ^{+ 0.008 }_{- 0.008 }$\\
 \hline
$H_{0}$[km s$^{-1}$ Mpc$^{-1}$]  & $ 73.142 ^{+ 1.315 }_{- 1.306 }$ & $ 73.864 ^{+ 1.285 }_{- 1.283 }$ & $\mathbf{74.101 ^{+ 1.124 }_{- 1.137 }}$ & $ 73.708 ^{+ 1.419 }_{- 1.413 }$\\  
 \hline
Goodness of fit & & & & \\
\hline
$\Delta\ln L_{\rm SN\,cal\,max}$ & 0 & 6.72 & $\mathbf{6.72}$ & 1.09 \\
 & & & & \\
$\Delta {\rm BIC}_{\rm SN\,cal}$ & 0 & $-7.55$ & $\mathbf{-10.49}$ & 0.77 \\
\end{tabular}
\caption{Best fit primary parameters of the SN data block measured from observations of Cepheids and type Ia SNe assuming the baseline model and its three extensions: model A (free $\beta_{\rm cal}$ and $\sigma_{\rm int\,cal}$), model B (free $\beta_{\rm cal}$ and $\sigma_{\rm int\,cal}\equiv0$) and model C (free $\beta_{\rm cal}$ and $\sigma_{\rm int\,cal}\equiv\sigma_{\rm int}$). Best fit results are provided in the form of the posterior mean values and errors containing 68 per cent of marginalised probabilities (with 68 percent upper limit for $\sigma_{\rm int\,cal}$ in model A). 
The Hubble constant for models A, B and C is derived assuming reference colour $c_{\rm ref}=0$ in eq.~(\ref{refcol}). In general, the Hubble constant determination depends on the choice 
of $c_{\rm ref}$ in these cases and the {\em Planck} value is obtained for $c_{\rm ref}\lesssim-0.13$, as discussed in subsection 3.3. Since extra parameters in the new models are constrained solely by the data from the SN calibration block, the goodness of fit is quantified in terms of the corresponding maximum likelihood $\Delta\ln L_{\rm SN\,cal\,max}=\ln L_{\rm SN\,cal\,max}{\rm (new\,model)}-\ln L_{\rm SN\,cal\,max}{\rm (baseline\,model)}$ and the Bayesian Information Criterion $\Delta{\rm BIC}=\Delta n\ln (19)-2\Delta\ln L_{\rm SN\,cal\,max}$, where $\Delta n=2,1,1$ for models A, B and C, respectively, and $N=19$ is the number of SNe in the SN calibration sample. Best fit parameters of the Cepheid data block are listed in Table~\ref{bestmodels-Ceph}.
}\label{bestmodels}
\end{center}
\end{table*}

We use the byproduct constraints on the distance moduli of the calibration galaxies ($\mu_{1}, ...,\mu_{19}$ in Table~\ref{table-likelihood}) to 
check if the distribution of residuals in the SN block confirm that the baseline model provides a complete and unbiased description of the data. Our 
study reveals what appears to be an anomalous relation between residual distance moduli of the SNe in the calibration sample and the SN colour parameter 
(see the left panel in Figure~\ref{19-res-c}). This trend suggests
that the universality of colour corrections assumed in the baseline model is inconsistent with the SN data. 
The apparent overestimation of the distances of red SNe ($c>0$) and the corresponding distance underestimation of blue SNe ($c<0$) 
appears to be comparable to the intrinsic scatter found in cosmological SN samples. This means that 
the anomaly can be concealed in the intrinsic scatter of the baseline model. This made it particularly difficult to detect in 
previous analyses. Fig.~\ref{19-res-c} also shows that that observed SN colours are distributed 
similarly in the calibration and cosmological samples, and are consistent with the corresponding colour distribution from the 
most recent SN compilation \citep[Pantheon+;][]{Brout2022}.

\begin{figure*}
	\centering
	\includegraphics[width=\linewidth]{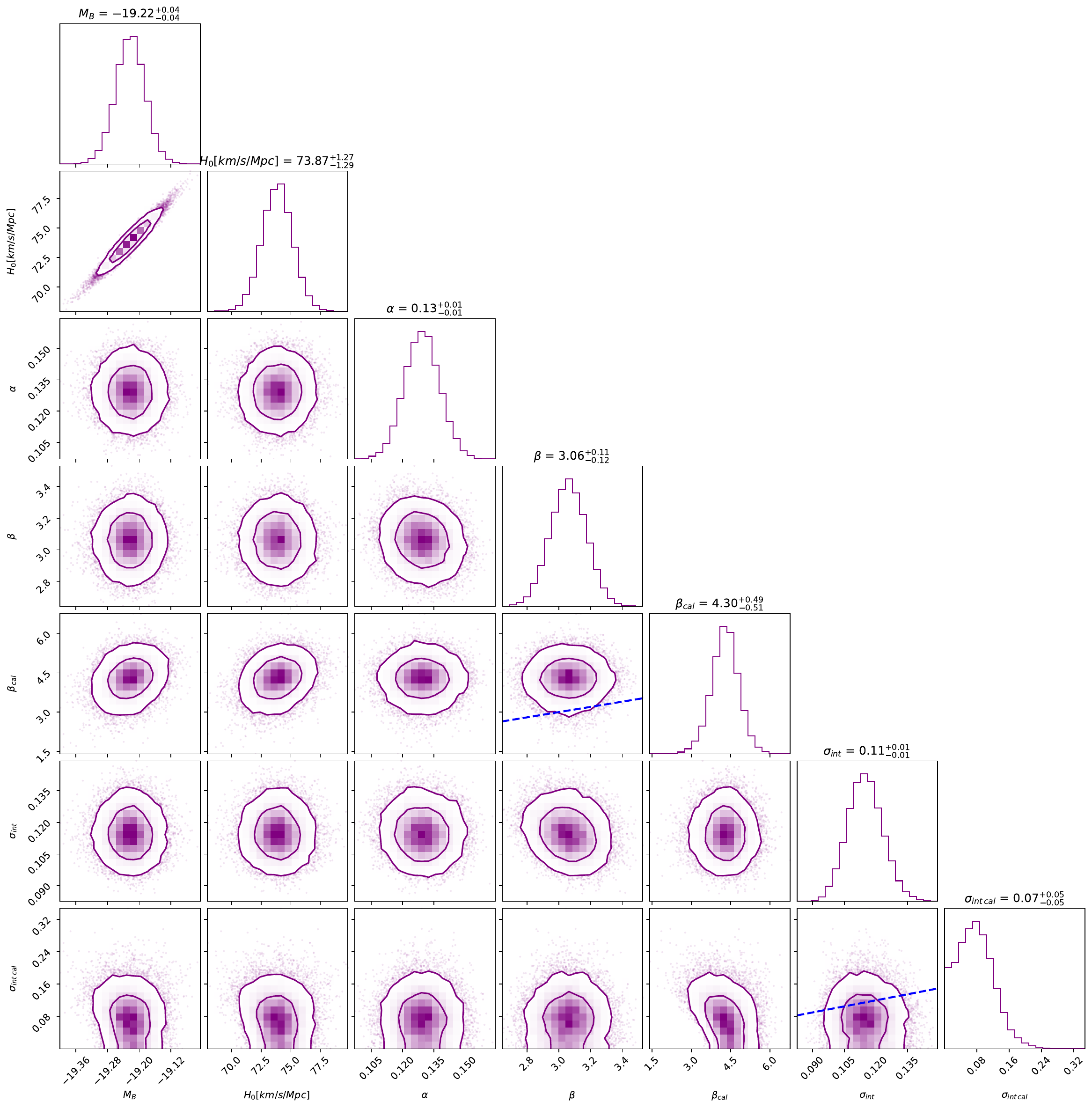}
	\caption{Constraints on parameters of the SN standardisation and the Hubble constant obtained for model A in which SN colour correction and intrinsic scatter are assumed to be independent in the calibration and cosmological samples. The blue dashed lines show 
	combinations of parameters which reduce model A to the baseline model which assumes a universal colour correction and intrinsic scatter across all SNe, i.e., 
	$\beta=\beta_{\rm cal}$ and $\sigma_{\rm int}=\sigma_{\rm int\,cal}$. The results demonstrate that SNe in the calibration 
	sample require stronger colour correction ($\beta_{\rm cal}>\beta$) than those in the cosmological sample. It is also apparent that the intrinsic 
	scatter in the calibration sample with an independent SN colour correction is consistent with 0, which is not the case of the baseline model. 
	The contours show 1$\sigma$ and $2\sigma$ confidence regions containing 68 and 95 per cent 
	of 2-dimensional marginalised probability distributions. The summary statistics above diagonal panels shows the median values and errors 
	given by 16-th and 84-th quantiles of the marginalised probability distributions.
	}
	\label{SN_params_2par_extension}
\end{figure*}

We quantify the statistical significance of the trend shown in Figure~\ref{19-res-c} by fitting a linear model. Keeping in mind that the apparent discrepancy 
between the SN colour correction in the calibration sample and the cosmological sample can also worsen the fit in the SN cosmological sample, although to a lesser extent, 
we expect that this approach gives us a lower limit of the actual significance of the anomaly. A complete and rigorous analysis is presented in the following 
subsection where we compare the baseline model to its minimum extensions motivated by the new trend found in the data. Taking into account uncertainties 
in both variables and assuming that they are uncorrelated, we find a positive correlation between residual distance moduli $\Delta\mu$ and SN colours $c$ 
(Fig.~\ref{19-res-c}) at the $\sim3\sigma$ significance level, with a linear slope of $1.10_{-0.30}^{+0.32}$. The slope is $0.93_{-0.42}^{+0.41}$ if intrinsic scatter is included as an extra free nuisance parameter.

\subsection{Extensions to the baseline model}

We now consider three models which allow for independent colour corrections in the calibration and cosmological samples. 
The intrinsic scatter in the calibration sample is assumed to be either an extra independent parameter (model A), 
vanishing ($\sigma_{\rm int\,cal}\equiv0$, model B), or equal to the analogous scatter in the cosmological sample 
($\sigma_{\rm int\,cal}\equiv\sigma_{\rm int}$, model C). We summarise the best fit SN parameters in Table~\ref{bestmodels}. All modifications with 
respect to the baseline model occur for parameters which are relevant for both SN data blocks, while Cepheid calibration parameters 
remain virtually unchanged and are provided in Table~\ref{bestmodels-Ceph} (appendix) for the sake of completeness. The constraints on SN parameters 
($M_{B},\alpha,\beta,\sigma_{\rm int},\beta_{\rm cal},\sigma_{\rm int\,cal}$) and the Hubble constant in model A  are shown in 
Figure~\ref{SN_params_2par_extension}. The red lines indicate combinations of parameters reducing model A to the baseline 
model, i.e., $\beta=\beta_{\rm cal}$ and $\sigma_{\rm int}=\sigma_{\rm int\,cal}$. The right panel of Fig.~\ref{19-res-c} demonstrates 
explicitly how the anomalous trend in distance modulus residuals obtained in the baseline model vanishes in model B (the most preferred by the data, as explained below).

\begin{figure}
	\centering
	\includegraphics[width=\linewidth]{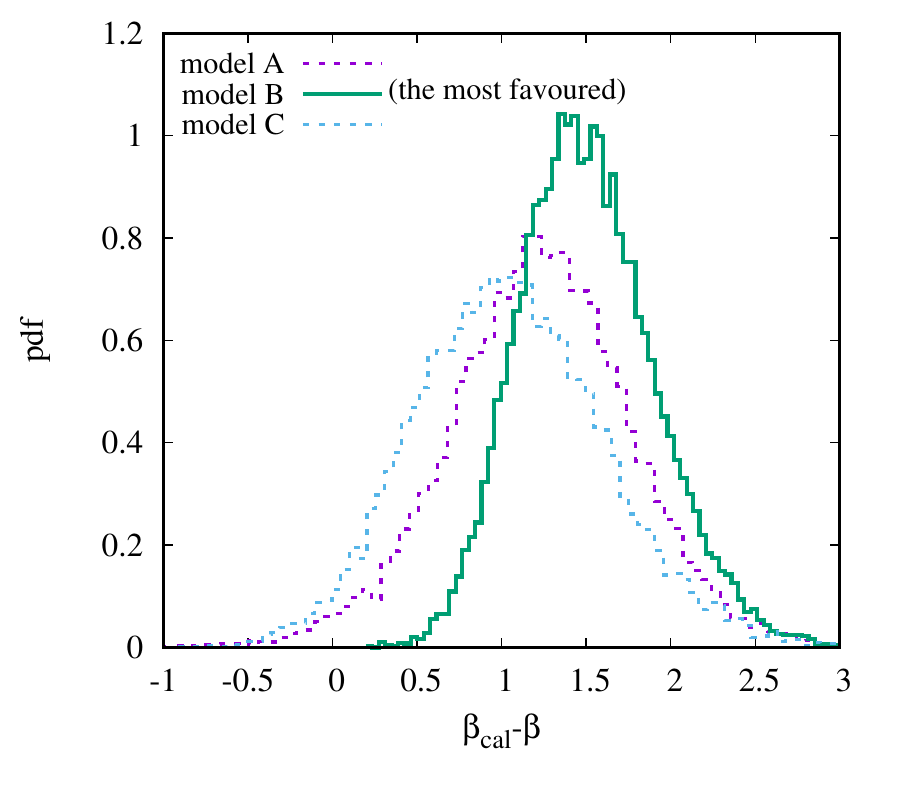}
	\caption{Discrepancy between the slope of the SN colour correction in the calibration ($\beta_{\rm cal}$) and 
	cosmological ($\beta$) samples. The curves show the probability distributions of the differences between the 
	colour correction slopes in the two SN samples obtained for three models: model A 
	($\sigma_{\rm int\,cal}$ free), model B ($\sigma_{\rm int\,cal}=0$) and model C ($\sigma_{\rm int}=\sigma_{\rm int\,cal}$). 
	Observations strongly favour a steeper slope of the colour correction in the calibration sample
	with $\beta_{\rm cal}=\beta+1.23\pm0.54$ (model A),  $\beta_{\rm cal}=\beta+1.51\pm0.40$ (model B) and 
	$\beta_{\rm cal}=\beta+1.05\pm0.56$ (model C). The largest difference between $\beta_{\rm cal}$ and $\beta$ is found for mode B which is the most favoured by the data.
	}
	\label{beta-diff}
\end{figure}

While the three models (A,B,C) treat the intrinsic scatter in the peak magnitudes in the calibration sample differently, they all
consistently show that the slope of the SN colour correction, $\beta_{\rm cal}$, in the calibration sample is larger than in the cosmological sample $\beta$. We find $\beta_{\rm cal}-\beta=1.23\pm0.54$ for model A, $\beta_{\rm cal}-\beta=1.51\pm0.40$ 
for model B and $\beta_{\rm cal}-\beta=1.05\pm0.56$ for model C. Figure~\ref{beta-diff} shows the marginalised posterior distribution for $\beta_{\rm cal}-\beta$ 
obtained for the three models. Numerical estimation of the probability that $\beta_{\rm cal}\leq\beta$ yields $0.017$ ($2.4\sigma$) for 
model A, $\sim10^{-4}$ ($3.8\sigma$) for model B and $0.025$ ($2.2\sigma$) for model C.

Table~\ref{bestmodels} provides goodness of fit quantified by the maximum likelihood 
and the Bayesian Information Criteria (BIC). Since all three extensions to the baseline model are constrained solely by the data in the SN calibration sample, we compute both metrics using $\ln L_{\rm SN\,cal}$ from the corresponding data block. 
We find strong evidence favouring model B ($\Delta{\rm BIC}\ll -6$) over the baseline model. Model A yields a substantially better fit than the baseline model, but its free intrinsic scatter $\sigma_{\rm int\,cal}$ turns out to be a redundant parameter (no detection: the maximum likelihood found at $\sigma_{\rm int\,cal}=0$ and fit yielding merely an upper bound, see Table~\ref{bestmodels} and Figure~\ref{SN_params_2par_extension}) and for this reason the model is less preferred than model B. The predictive power of model C is substantially reduced by the fact that a large part of the anomaly is effectively absorbed by intrinsic scatter which is fixed at the value inferred from the cosmological sample. Consequently, model C is less favoured than the baseline model in terms of BIC, although fully consistent with the data.

The strongest discrepancy between $\beta_{\rm cal}$ and $\beta$ is found for model B which is the most favoured by the data among the
three extensions to the baseline model considered here. The vanishing intrinsic scatter in the calibration sample assumed 
in this model indicates that independent colour correction results in maximally improved precision of the SN standardisation in the calibration sample. 
This can be also realised 
by fitting a model with a universal colour correction ($\beta=\beta_{\rm cal}$) and free $\sigma_{\rm int\,cal}$. The fit 
yields $\sigma_{\rm int\,cal}=0.135\pm0.034$ demonstrating directly that the assumption of universal SN colour correction substantially degrades the precision of the SN standardisation in the calibration sample. It is also worth emphasizing that model B has a greater predictive power than the baseline model in a broader context. As we discuss in the following section, the new SN colour correction in the calibration sample can be straightforwardly linked to the standard dust extinction known from the Milky Way.

\subsection{Hubble constant}

When defining the SN absolute magnitude $M_{\rm B}$ and the corresponding standardised magnitudes we have the freedom to choose an arbitrary reference colour parameter $c_{\rm ref}$, i.e.,
\begin{eqnarray}
&m_{B}=\mu+M_{B}+\alpha x_{1}+\beta(c-c_{\rm ref})&{\rm for\;SN\;block,}\nonumber \\
&m_{B}=\mu+M_{B}+\alpha x_{1}+\beta_{\rm cal}(c-c_{\rm ref})&{\rm for\;SN\;cal\;block}.\nonumber \\
\label{refcol}
\end{eqnarray}
Changing $c_{\rm ref}$ in the baseline model ($\beta=\beta_{\rm cal}$) automatically modifies $M_{B}$ inferred from the data, but it does not affect the remaining parameters, including $H_{0}$. From this point of view, the commonly adopted $c_{\rm ref}=0$ is an arbitrary 
choice which does not have any implications for the Hubble constant determination and can be merely motivated by the
minimisation of a correlation between $M_{\rm B}$ and $\beta$. However, this situation changes when the colour 
correction is not universal and differs between the calibration and cosmological SN samples, i.e., $\beta\neq\beta_{\rm cal}$. In this case, 
adopting $c_{\rm ref}\neq 0$ changes the SN absolute magnitude (with respect to the $c_{\rm ref}=0$ case) differently in the calibration and cosmological samples, by $\Delta M_{B}=\beta_{\rm cal}c_{\rm ref}$ and $\Delta M_{B}=\beta c_{\rm ref}$, respectively. The only way to reconcile SN absolute magnitudes in 
both SN samples is to adjust the Hubble constant so that the difference in $\Delta M_{B}$ is fully compensated by the corresponding shift in distance moduli $\Delta\mu$:
\begin{equation}
    \Delta\mu=(\beta-\beta_{\rm cal})c_{\rm ref}.
    \label{deltamu}
\end{equation}
This means that every choice of $c_{\rm ref}$ has its unique best fit value of the Hubble constant inferred from the same observational data.

The values of the Hubble constant measured for models A, B, C and listed in Table~\ref{bestmodels} are obtained for a SN 
absolute magnitude defined for the same reference colour as in the baseline model ($c_{\rm ref}=0$). Since the reference 
colour is very close to the median colour in both SN samples, the resulting constraints on the Hubble constant agree 
fairly well with the result from the baseline model. For small reference colours, i.e., $|c_{\rm ref}|<1$, we derive from eq.~(\ref{deltamu}) that the best fit Hubble constant shifts 
by $\Delta H_{0}$ with respect to its value obtained for $c_{\rm ref}=0$ as given by the following equation:
\begin{equation}
\Delta H_{0}/H_{0}\approx (\ln 10/5)(\beta_{\rm cal}-\beta)c_{\rm ref}.
\end{equation}
Using the above approximation we can show that models A, B or C can recover the {\em Planck} value of the Hubble constant 
respectively for $c_{\rm ref}\approx-0.16,-0.13, -0.18$. The reference colours required to obtain the {\em Planck} value of the Hubble constant 
coincide with the blue end of the observed colour distribution in the cosmological and calibration sample. Interestingly, the reference colour for the preferred model B is within the $1\sigma$ range of the intrinsic colour distribution derived from the Pantheon+ SN sample assuming an exponential distribution of dust reddening \citep{Popovic2021}.

\begin{figure}
	\centering
	\includegraphics[width=\linewidth]{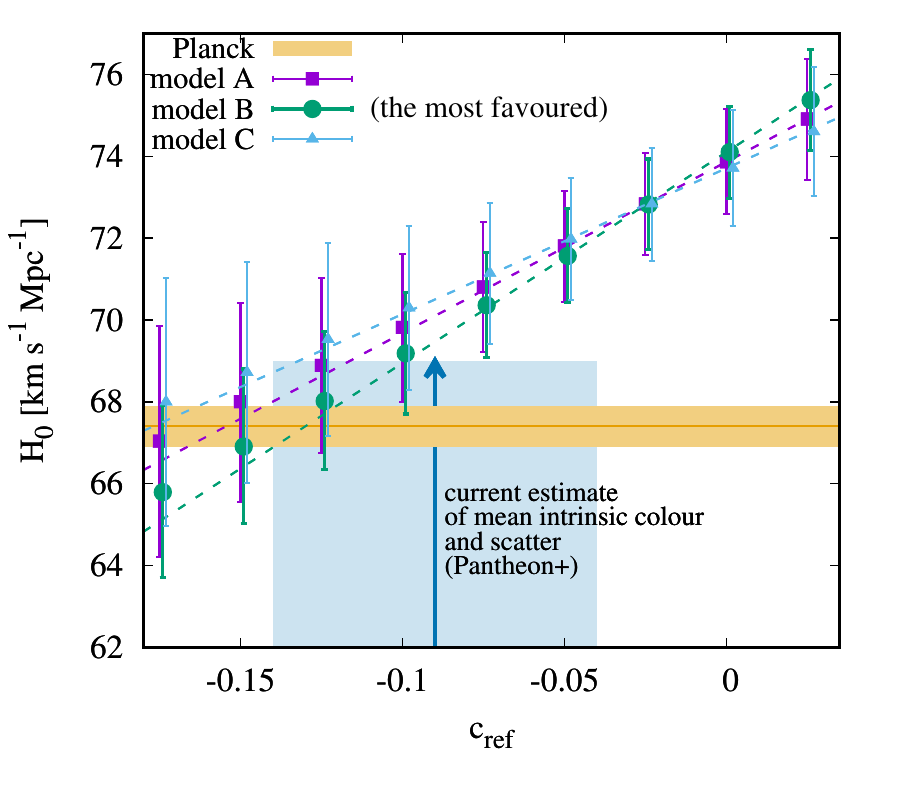}
	\caption{The Hubble constant determined from type Ia SN observations calibrated with Cepheids as 
	a function of reference colour $c_{\rm ref}$ (see eq.~\ref{refcol}) in models enabling different slopes of the SN 
	colour correction in the calibration and cosmological samples. The results from the three models  (see Table~\ref{bestmodels} for 
	differences) obtained for the same reference colour are slightly shifted horizontally for the sake of readability. The yellow line and the 
	corresponding shaded region show the {\em Planck} measurement obtained for the standard flat $\Lambda$CDM cosmological 
	model. The results show that the new local $H_{0}$ measurement decreases the Hubble tension to a sub-2$\sigma$ confidence level 
	for $c_{\rm ref}\approx -0.06$ and recovers the {\em Planck} measurement for $c_{\rm ref}\approx-0.13$ (for the most preferred model B). The blue arrow and the shaded rectangle mark the current estimate 
	of the mean SN intrinsic colour and scatter derived from Pantheon+ SN sample assuming an exponential distribution of dust reddening \citep{Popovic2021}.
		}
	\label{H0-c-ref}
\end{figure}

Following the above estimates of $c_{\rm ref}$ required for a reduction of the Hubble constant to the {\em Planck} value, we repeat the
full analysis 
with models A, B and C for a range of $c_{\rm ref}$ values between $-0.175$ and $0.025$. The only modification of the 
corresponding likelihoods listed in Table~\ref{table-likelihood} occurs in the two SN blocks where the reference colour modifies 
SN distance moduli according to eq.~(\ref{refcol}). Figure~\ref{H0-c-ref} shows the resulting constraints on the 
Hubble constant as a function of the reference colour. We find that all three models reduce the 
Hubble constant tension to sub-2$\sigma$ confidence levels when $c_{\rm ref}\approx-0.06$ and recover the {\em Planck} measurement 
when $c_{\rm ref}\approx-0.15$. The strongest modification of the best fit $H_{0}$ as a function of $c_{\rm ref}$ occurs 
for model B ($\sigma_{\rm int\,cal}=0$) which yields the largest difference between the colour correction slopes in the calibration sample
and cosmological sample. The apparent reduction of the Hubble constant tension appears primarily due to gradually decreasing 
best fit $H_{0}$ and to a much lesser extent due to larger errors. The latter results from a degeneracy between the colour correction slope and 
$H_{0}$. This degeneracy is minimised at reference colour $c_{\rm ref}\approx-0.02$, at which a correlation between $\beta_{\rm cal}$ 
and $M_{B}$ happens to vanish for this particular SN calibration sample, but it gradually increases for $c_{\rm ref}<-0.02$ or $c_{\rm ref}>-0.02$. The 
effect of vanishing correlation between $\beta_{\rm cal}$ and $M_{B}$ is visible in Fig.~\ref{H0-c-ref} as the smallest errors and differences between best 
fit $H_{0}$ values at $c_{\rm ref}\approx-0.02$. We find that the errors in $H_{0}$ measurements for $c_{\rm ref}=-0.125$ are $\sim60$ per cent larger than for $c_{\rm ref}=0$.

\section{Discussion}

Our analysis shows that the assumption of a universal colour correction expressed in eq.~(\ref{SN-cal}) does not hold for the SN 
sample used in the local determination of the Hubble constant. Colour correction in the calibration sample 
requires a steeper slope than for the remaining SNe in the Hubble flow. A direct implication of this finding is that the Hubble 
constant determination becomes dependent on the choice of a reference colour, with the {\em Planck} value recovered for $c_{\rm ref}\approx-0.13$ (model B). This new property is entirely inferred from the observational data.

As shown in Fig.~\ref{H0-c-ref}, the Hubble constant tension decreases to just $1\sigma$ when the adopted reference colour in eq.~(\ref{refcol}) is the mean intrinsic colour of type Ia SNe, as measured recently from the Pantheon+ sample \citep{Popovic2021}. 
It is also important to notice that the relative colour $c-c_{\rm ref}$ in the calibration sample becomes strictly positive (see Fig.~\ref{19-res-c}). 
With this mind, it is natural to suspect that the colour correction in this case is predominantly driven by dust extinction. This picture is consistent 
with the probabilistic model of SN colours used by \citet{Popovic2021} for which the red tail of the observed colour distribution arises entirely 
from reddening due to dust. Following this interpretation, we conclude that a low, nearly Planckian value of the Hubble constant (or a longer distance scale) 
obtained in our analysis for $c_{\rm ref}\lesssim -0.1$ results most likely from a stronger dust extinction and thus higher intrinsic brightness ($\Delta m\approx0.17$ for $c_{\rm ref}=-0.13$ and model B) of SNe in the calibration sample than in the baseline model.

Our analysis recovers the {\em Planck} value of the Hubble constant for reference colour $c_{\rm ref}\approx-0.13$ (the most favoured model B) which happens 
to coincide with the bluest colour of SNe in both the calibration and cosmological samples. This brings us to a proposal in which 
full agreement with the {\em Planck} cosmology can be restored when we assume that intrinsic SN colours is close to $-0.13$ 
(with negligible scatter), while redder SN colours result simply from reddening by intervening dust. The required intrinsic colour can be obtained from SN observations following a similar approach to that presented in \citet{Popovic2021,Brout2021}, but with a modified model describing the distribution of colour excess, $E(B-V)$. \citet{Popovic2021} employs an exponential distribution of $E(B-V)$ proposed by \citet{Mandel2011}. This model can 
be justified as a maximum entropy solution given the mean value of the colour excess. However, this is only a motivation based on 
information theory which does not necessarily reflect any physical constraints such as the 3D distribution of dust and SNe in galaxies. In fact, 
the distribution of apparent colours seen in Fig.~\ref{19-res-c} can be reproduced by a wider range of combinations of intrinsic colour 
and reddening distributions. In particular, the simplest solution motivated by minimising the Hubble constant tension is a single-valued 
intrinsic colour $c_{\rm int}\approx-0.13$ (or a very narrow distribution with the mean value equal to $c_{\rm int}$) and a colour excess 
distribution given by $c-c_{\rm int}$. We think that this scenario is worth further exploration as a possible revision of the exponential distribution 
of dust reddening assumed in the present Bayesian models \citep[see e.g.][]{Mandel2017,Thorp2021}. One should also 
keep in mind that the intrinsic colour of a SN may depend on other intrinsic properties \citep[e.g.][]{Foley2011}.

Assuming that the colour excess with respect to reference colour $c_{\rm ref}\lesssim -0.13$ arises predominantly as dust reddening, i.e., 
$c-c_{\rm ref}=E(B-V)$, we can interpret $\beta$ (or $\beta_{\rm cal}$) in our analysis with models A, B or C as the extinction coefficient $R_{B}$. In this picture, 
galaxies in the calibration sample appear to be analogs of the Milky Way in terms of their dust properties. The extinction coefficient in the calibration sample 
determined in our analysis ($R_{B}\approx4.6\pm0.4$ for model B in Table~\ref{bestmodels}) is fully consistent with the mean extinction coefficient measured in the Milky Way, i.e., $R_{B}\approx 4.3$ \citep{Fitzpatrick1999,Cardelli1989}. Furthermore, Milky-Way-like extinction in the calibration sample readily improves the precision of the SN 
standardisation decreasing the intrinsic scatter in distance moduli from $\sigma_{\rm int\,cal}=0.135\pm0.034$ to $\sigma_{\rm int\,cal}<0.097$. 
Finally, the measured extinction coefficient in the calibration sample ensures full consistency with the extinction correction applied to Cepheid 
observations which is based on $R_{\rm B}=4.3$ \citep{Riess2019,Riess2021}.

The above dust-based interpretation of SN colours is less straightforward in the cosmological sample. SN host galaxies in the Hubble flow 
are characterised by a substantially lower coefficient of the colour correction with $\beta=3.06\pm0.11$. This result is consistent with many other studies of various 
SN cosmological sample \citep[see e.g.][]{Jones2019,Scolnic2018} and it has been a long lasting problem whether it can be interpreted as extinction using physical models of dust. Although Bayesian analysis of type Ia SNe suggest that a broad 
distribution of $R_{B}$ spanning between $R_{B}\approx 2$ and $R_{B}\approx5$ is favoured by SN data 
\citep[as a means to reduce an appreciable fraction of scatter in the Hubble diagram; see][]{Brout2021,Popovic2021}, a complete physical picture connecting dust properties and extinction inferred from observations is still missing. In particular, the extinction coefficient measured in the 
Milky Way falls into a rather wide range $3.5\lesssim R_{B}\lesssim 6.5$ \citep{Cardelli1989} 
while values of 2--3 are not seen.
The problem becomes particularly pressing in the context of our study in which the apparent differences between the extinction properties in galaxies of the 
calibration and cosmological samples 
can be used to reconcile the local and CMB-based measurements of the Hubble constant.

\section{Summary and Conclusions}

We have reanalysed Cepheid and type Ia SN observations used in the most precise local determination of the Hubble constant 
to date to test the universality of the commonly used phenomenological colour correction for
SN standardisation given by eq.~(\ref{SN}) proposed by \citet{Tripp1998}. Our main results are:

(i) SN data in the calibration sample (galaxies with observed Cepheids) and cosmological sample (galaxies in the Hubble flow) 
are inconsistent with the assumption of a universal colour correction. Standardisation of SNe in the calibration sample requires 
a steeper slope found at $2.2\sigma$ confidence level assuming universal intrinsic scatter, 
$2.4\sigma$ confidence level when fitting intrinsic scatter independent in the calibration sample 
and at $3.8\sigma$ confidence level assuming 
vanishing intrinsic scatter in the calibration sample. The latter model maximising the difference between the SN colour corrections in the two SN samples is the most favoured by the data. Accounting for the SN colour correction inferred directly from the SN calibration data eliminates the necessity for including intrinsic scatter in the corrected SN peak magnitudes.

(ii) The difference between SN colour corrections in the calibration and cosmological samples inevitably makes the Hubble constant 
measurement dependent on the choice of reference colour setting the absolute magnitude in both SN samples. While the Hubble 
constant value of \citet{Riess2022} is recovered for $c_{\rm ref}=0$, we find that gradually lower values are measured 
when using gradually bluer reference colour, i.e. $c_{\rm ref}<0$. We recover the {\em Planck} value for $c_{\rm ref}\approx-0.13$ 
which happens to coincide with the blue end of the apparent colour distribution.

(iii) The slope of the SN colour correction in the calibration sample coincides numerically with the mean extinction coefficient found in 
the Milky Way. This suggests that galaxies in the calibration sample -- unlike SN host galaxies in the cosmological sample -- are 
analogs to the Milky Way in terms of their dust extinction properties.

(iv) The minimum physical scenario required to obtain the {\em Planck} value of the Hubble constant assumes that the SN intrinsic colour is $-0.13$ (with relatively small scatter) and the observed colour distribution results predominantly from dust reddening with Milky-Way-like extinction in the calibration sample.

Our study opens up a new avenue for understanding the physical origin of the Hubble constant tension. In this proposal, the tension 
arises from insufficiently accurate standardisation of type Ia SNe resulting from a poor understanding of dust extinction and 
SN intrinsic colours. Further investigation and perhaps more observations will be needed to test this scenario and address 
arising questions. Analysis of alternative calibration samples based on TRGB or SBF distance calibrations will allow us to 
verify if the new colour correction shown in our study is a special property of host galaxies with well observed Cepheids or a 
generic property of SN host galaxies 
in the nearby universe. Additional tests can be enabled by spectroscopic observations of the local SN environments which provide independent constraints on 
dust content along the lines of sight to SNe. In general, the empirically 
determined SN colour correction can result from two physical effects: reddening due to dust and/or a possible modulation related to SN intrinsic colour. Disentangling these effects will be essential for understanding the colour correction from first principles. This will perhaps require a better synergy between advanced data analysis methods 
and physically motivated models of dust extinction and light emission by type Ia SNe, following the recent studies by  \citet{Mandel2017,Thorp2021,Popovic2021}. Another avenue worth pursuing are observations of SNe in near infrared which minimize the effect of dust extinction. Recent studies reported the Hubble constant estimates consistent with the SH0ES value \citep{Suhail2018,Burns2018,Jones2022}. However, more effort is needed to reduce the measurement errors which are currently too large to confirm conclusively the Hubble constant tension.

\section*{Acknowledgments}
This work was supported by a VILLUM FONDEN Investigator grant (project number 16599). RW thanks Adriano Agnello, Charlotte Angus, Christa Gall, Darach Watson, Luca Izzo and Nandita Khetan for discussions, and David Jones for constructive comments. The authors thank the anonymous referee for constructive comments which helped improve this work.

\section*{Data availability}
No new data were generated or analysed in support of this research.

\bibliography{master}

\appendix
\onecolumn

\section{Best fit parameters of the Cepheid data}

\begin{table*}
\begin{center}
\begin{tabular}{lcccc}
\hline
 & baseline & model A & \textbf{model B}  & model C \\ 
 & & & \textbf{(most favoured)} & \\
  \hline
 Cepheid parameters & & & & \\
 \hline
$M_{\rm F160W}$ & $ -5.929 ^{+ 0.014 }_{- 0.014 }$ & $ -5.929 ^{+ 0.014 }_{- 0.014 }$ & $\mathbf{-5.929 ^{+ 0.014 }_{- 0.014 }}$ & $ -5.929 ^{+ 0.014 }_{- 0.014 }$\\
 &  & & & \\
$b_{W}$ & $ -3.292 ^{+ 0.012 }_{- 0.012 }$ & $ -3.292 ^{+ 0.012 }_{- 0.012 }$ & $\mathbf{-3.292 ^{+ 0.012 }_{- 0.012 }}$ & $ -3.292 ^{+ 0.012 }_{- 0.012 }$ \\
 &  & & & \\
$z_{W}$ & $ -0.213 ^{+ 0.049 }_{- 0.048 }$ & $ -0.209 ^{+ 0.047 }_{- 0.047 }$ & $\mathbf{-0.208 ^{+ 0.046 }_{- 0.046 }}$ & $ -0.211 ^{+ 0.048 }_{- 0.048 }$ \\
 & & & & \\
$zp$[$\mu$as]  & $ -17 ^{+ 4 }_{- 4 }$ & $ -17 ^{+ 4 }_{- 4 }$ & $\mathbf{ -17 ^{+ 4 }_{- 4 }}$ & $ -17 ^{+4 }_{-4 }$\\
\end{tabular}
\caption{Best fit parameters of the Cepheid data block measured from observations of Cepheids and type Ia SNe assuming the baseline model and its three extensions: model A (free $\beta_{\rm cal}$ and $\sigma_{\rm int\,cal}$), model B (free $\beta_{\rm cal}$ and $\sigma_{\rm int\,cal}\equiv0$) and model C (free $\beta_{\rm cal}$ and $\sigma_{\rm int\,cal}\equiv\sigma_{\rm int}$). Best fit results are provided in the form of the posterior mean values and errors containing 68 per cent of marginalised probabilities.
}\label{bestmodels-Ceph}
\end{center}
\end{table*}

\end{document}